\documentclass[twocolumn,showpacs,preprintnumbers,amsmath,amssymb]{revtex4}

\usepackage{graphicx}
\usepackage{dcolumn}
\usepackage{bm}

\begin{document}

\preprint{APS/123-QED}

\topmargin 0pt

\title{Co-community Structure in Time-Varying Networks} 

\author{Shihua Zhang}
\email{zsh@amss.ac.cn}
\author{Junfei Zhao}
\author{Xiang-Sun Zhang}

\affiliation{%
$^1$National Center for Mathematics and Interdisciplinary Sciences,
Academy of Mathematics and Systems Science, Chinese Academy of
Sciences, Beijing 100190, China}

\date{\today}

\begin{abstract}
In this report, we introduce the concept of co-community structure
in time-varying networks. We propose a novel optimization algorithm
to rapidly detect co-community structure in these networks. Both
theoretical and numerical results show that the proposed method not
only can resolve detailed co-communities, but also can effectively
identify the dynamical phenomena in these networks.
\end{abstract}

\pacs{Valid PACS appear here}


\maketitle

Networks consisting of vertices and edges connecting some pairs of
vertices are powerful abstractions of relational data, hence have
become very popular tools in many fields including sociology,
biology and physics \cite{Freeman}. The characteristic of community
structure in networks, i.e., networks are naturally divided into
modules or communities, has attracted huge attention in the past
decade which can provide insights into the structure and dynamic
formation of networks. Many methods for community detection in one
network have been developed and studied even including the fuzzy
community structure identification problem \cite{Reichardt04} and
the more challenging community detection problem in directed
networks \cite{Leicht08} (see Ref. \cite{Newman04} for recent
comprehensive reviews).

However, previous studies have concentrated on uncovering community
structure in a static network which only represents a summarized
picture of all possible relations. A typical example is the protein
interaction network in biology which represent all proteins of an
organism and all interactions regardless of the conditions and time
under which interactions may take place \cite{Rachlin2006}. In
reality, most of relationships modeled by networks evolve with time
or conditions \cite{Dorogovtsev2002}.

Several recent studies have touched on the analysis of dynamic
networks including analyzing changes of global properties, detecting
anomalous changes, mining dynamic frequent subnets, and discovering
similar evolving regions in evolving networks \cite{Chan08} and even
the dynamic communities by combining the information of communities
in each network using traditional community detection methods.
However, the community structure in two or more slices of a series
of time-varying networks has not been well addressed directly
\cite{Mucha10,Rosvall10}.

In this report, we propose the concept of co-community structure in
two or more networks of a series of time-varying networks. The basic
assumption is that an essential and common community structure may
exist in two or more networks, and local dynamic changes may happen.
This is very realistic in time-varying networks of many robust
systems.

Suppose that we are given the structure of two or more networks of
the same vertices, we aim to determine whether there exists any
co-community structure, or say similar groups or communities in
these networks. Moreover, along this goal, we attempt to uncover the
dynamic characteristics of some vertices. Mathematically, the
co-community structure and dynamical characteristic are stored in
matrices which can be determined by an efficient optimization
procedure.

Let us focus initially on the problem in two networks that will be
more useful in analyzing time-varying networks. To formulate the
problem easily, we consider the common notation of clustering or
community structure detection problems. The objective of classical
community detection in networks is to partition the vertex set $V$
of the graph $G(V,E)$ with $|V|=N$ into $K$ distinct subsets in a
way that puts densely connected groups of vertices in the same
community. In this case, a convenient representation of a given
partition is the partition matrix $U = [u_{ik}]$ (or $[u_{i}]$,
$u_i$ is a membership vector) with size of $N\times K$
\cite{Nepusz08}. And $u_{ik} = 1$ if and only if vertex $i$ belongs
to the $k$th subset in the partition, otherwise it is zero. From the
definition of the partition, it clearly follows that $\sum_{k=1}^K
u_{ik} = 1$ for all $i$. The generalization of the hard partition
follows by allowing $u_{ik}$ to attain any real value from the
interval $[0, 1]$, and the corresponding matrix is also called
membership matrix.

In the following, we adopt the popular membership matrix
representation to formulate the problem. Like Nepusz \emph{et al.}
\cite{Nepusz08} have suggested that an edge between vertex $v_1$ and
$v_2$ implies the similarity of $v_1$ and $v_2$, and likewise, the
absence of an edge implies dissimilarity, i.e, $a_{ij}\simeq u_i
u_j^T$ or $A\simeq U U^T$, where $A =(a_{ij})$ is the adjacency
matrix of a network. At the same time, the same vertices in two
networks should have similar membership vectors. These
considerations can be formulated as:
\begin{equation}\label {eq11}
\begin{array}{l}\min
\sum\limits_{g=1}^2 \|A_g-H_gH_g^T\|_F^2+\lambda_1
\sum\limits_{g=1}^2 \|H_g-H\|_1 + \lambda_2 \|H\|_1
\end{array}
\end{equation}
\begin{equation*}
\left. s.t.\{ \quad \begin{array}{l}
   \sum_{k=1}^K (H_g)_{ik} = 1; (H_g)_{ik}, H_{ik}\geq 0;\\
    g=1,2, i=1,\cdots,N, k=1,\cdots,K.
\end{array}
\right.
\end{equation*}
where $A_g$ is the adjacent matrix of network $G(V,E_g)$, $H_g$ is
the membership matrix of network $G(V,E_g)$, $H$ is the virtual
co-membership matrix representing the membership of nodes reflected
in all networks, $\|\cdot\|_F$ and $\|\cdot\|_1$ are the entrywise
matrix norm ($\|\cdot\|_F$ is known as the Frobenius norm). To solve
the problem easily, we remove the constraints $\sum_{k=1}^K
(H_g)_{ik} = 1$ ($g=1,2$; $i=1,\cdots,N$). Then the magnitude of
$(H_g)_{ik}$ reflect the intensity of vertex $i$ belonging to the
community $k$ in the network $G(V,E_g)$. This formulation allows us
to map the communities of two networks as well as their
co-communities.

The non-convexity and the non-smoothness of the objective function
of Eq.(1) make it a more challenging mathematical programming
problem. To practically solve the problem (Eq.[1]), we employ a
decomposition technique. We can easily find that, given the
co-communities matrix $H$, the technique leads to two symmetrical
non-negative factorization matrix (SNMF) problems \cite{Ding05}
coupled with a penalty term as follows:
\begin{equation}\label {eq11}
\begin{array}{l}\min
\sum\limits_{g=1}^2 \|A_g-H_gH_g^T\|_F^2+\lambda_1
\sum\limits_{g=1}^2 \|H_g-H\|_1.
\end{array}
\end{equation}
Fortunately, it can be divided into two independent subproblems
which can be solved in a symmetric NMF manner with the following
updating rule:
\begin{equation}\label {eq11}
\begin{array}{l}
(H_g)_{ik}\leftarrow
(\widetilde{H_g})_{ik}\left(1-\beta+\beta\frac{(A_g\widetilde{H_g})_{ik}}{(\widetilde{H_g}\widetilde{H_g}^T\widetilde{H_g})_{ik}}\right),
\end{array}
\end{equation}
where $\widetilde{H_g}=H_g+\Delta(H_g-H)$, and $0<\beta\leq 1$ (we
find $\beta=1/2$ is a good choice). The first term of Eq. (2) may
dominant the optimization procedure, then the columns of the two
decomposition matrices may be inconsistent in terms of their
membership profiles. So we reorder their columns by maximizing their
corresponding correlations to facilitate the optimization procedure.

While given the community matrix $H_g$ of each network, it leads to
the following problem:
\begin{equation}\label {eq11}
\begin{array}{l}\min
\lambda_1 \sum\limits_{g=1}^2 \|H_g-H\|_1 + \lambda_2 \|H\|_1.
\end{array}
\end{equation}
This formulation with positive combination of $L_1$ norm of
variables, can be transformed into a large-scale linear programming
problem through a well-known procedure. More interestingly, it can
be solved efficiently by a further decomposition technique
\cite{DT}. We should note, owing to $L_1$ norm, generally the
optimal solution has an excellent property, i.e., there are as many
zeros for $\|H_g-H\|_1$ and $\|H\|_1$ as possible. This point
exactly serves the final goal, i.e., consistency and sparseness of
the membership of each vertex.
\begin{figure}[t]
\begin{center}
\includegraphics[width=0.48\textwidth]{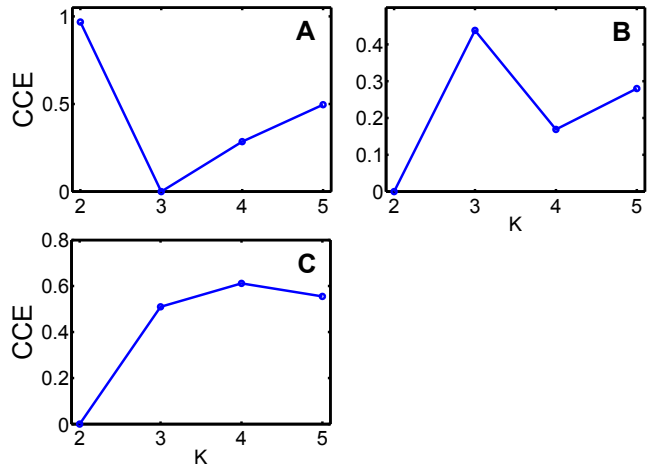}
\caption{The co-community entropy for each testing network system in
the following analysis: (A) The simulated networks; (B) The karate
club networks; (C) The U.S. senate networks.} \label{figurecurve}
\end{center}
\end{figure}

Therefore, we have the following algorithm for discovering
co-communities in two undirected networks. We first set the
parameters $\lambda_1$, $\lambda_2$, $\beta$ and $K$; and initialize
the membership matrices $H_1$and $H_2$, and set $H=H_1+H_2$. For the
subproblem Eq.(2), we use the update rule Eq.(3) to update $H_1$ and
$H_2$ respectively. Then using the new $H_1$ and $H_2$ we solve the
subproblem Eq.(4) to obtain the new $H$, by subdividing it into
$N\times K$ one-dimensional optimization subproblem. We iteratively
solve the subproblem Eq.(2) and Eq.(4) until $H$ doesn't change too
much (e.g., $\frac{\|H_{new}-H_{old}\|_F^2}{\|H_{old}\|_F^2}
<10^{-5}$, where $H_{new}$ and $H_{old}$ are the $H$ in current step
and last step respectively). The final $H$, $H_1$ and $H_2$ store
the co-communities and dynamical information. The $H$ ($H_1$ and
$H_2$) can be considered as a fuzzy partition of the network(s)
directly \cite{Zhang07}. It can also be employed to determine a hard
partition by assigning a node into a single community according to
the maximum value in each row of $H$ ($H_1$ and $H_2$)
\cite{Brunet04}.

The time complexity of the proposed algorithm is $O(TKN^2)$, where
$T$ is the number of iterations. The efficiency of the method can
also be seen in its application to networks with size of 10000 (see
Appendix). Note that the method can be applied onto a single network
by minimizing the criterion: $\| A_g-H_g H_g^T \|_F^2$ and it shows
competitive performance with two popular algorithms (see Appendix).

The formulation for two networks can be easily extended to more than
two networks as follows:
\begin{equation}\label {eq11}
\begin{array}{l}\min
\sum\limits_{g=1}^G \|A_g-H_gH_g^T\|_F^2+\lambda_1
\sum\limits_{g=1}^G \|H_g-H\|_1 + \lambda_2 \|H\|_1,
\end{array}
\end{equation}
where all the $H_g$ and $H$ are non-negative matrices. The algorithm
can also be easily extended.

The key issue in community detection is the proper choice of $K$.
Here, we employ the stochastic nature of the proposed algorithm to
achieve this. We should note that a similar strategy has been used
to determine the number of clusters in gene expression studies
\cite{Brunet04}. The differences and similarities of these
realizations is employed to evaluate the robustness of a partition
of given $K$. Specially, for each run, the vertices assignment can
be defined by a connectivity matrix $C$ of size $N\times N$, with
entry $c_{ij}$ if vertices $i$ and $j$ belong to the same
communities, and $c_{ij}=0$ if they belong to different clusters. We
can then compute the consensus matrix, $\overline{C}$, defined as
the average connectivity matrix over many runs. The entries of
$\overline{C}$ range from 0 to 1 and reflect the probability that
vertices $i$ and $j$ belong to one community.

From a more global point of view, we adopt the entropy as a measure
of the stability of the co-community structure. We assume that the
$c_{ij}$ are independent of each
other and we define the average Co-Community Entropy (CCE) score 
as:
$$
CCE = -\frac{2}{N(N-1)}\sum_{(i;j)} [c_{ij} \mbox{log}_2 c_{ij} + (1
- c_{ij}) \mbox{log}_2(1-c_{ij}],
$$
where the sum is taken over all edges and $m$ is the total number of
edges in the network. If the network is totally unstable (i.e., in
the most extreme case $c_{ij} = 0.5$ for all pairs), CCE $= 1$,
while if the edges are perfectly stable under noise ($c_{ij} = 0$ or
1), CCE $ = 0$. We have demonstrated that the CCE score can help to
select the number of communities in the time-varying networks
(Figure 1). For example, the CCE sore for the simulated networks
corresponds to very small value for $K=3$ which indicate that the
system have three distinct communities. We should note that the
parameters $\lambda_1$, $\lambda_2$ and $\beta$ can also be
evaluated with the CCE score by running the method with many trials.
\begin{figure}[t]
\begin{center}
\includegraphics[width=0.45\textwidth]{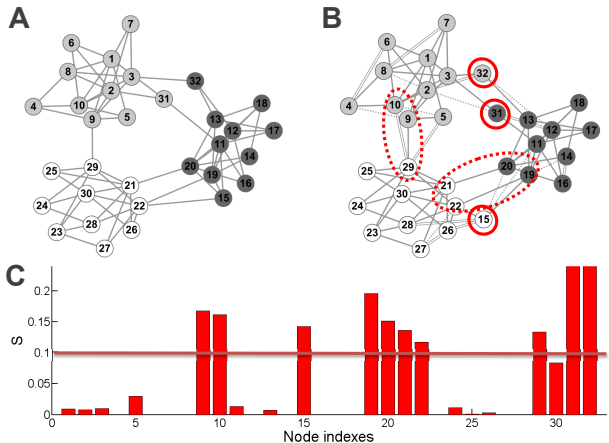}
\caption{Illustration of a toy example to show the major idea. (A)
The system under the first condition where the links were marked
with solid lines; (B) The system under the second condition with
links of some vertices changing, where dotted lines mean links exist
in previous condition, but disappear in current condition; while
double lines mean new links. (C) The dynamic index shows the dynamic
properties of vertices, of which with high values affecting the
community structure. The horizontal line was drawn to indicate
several distinct $S$ values, whose corresponding nodes have been
marked in (B). Similar line has been drawn in Figure 3C.}
\label{figurecurve}
\end{center}
\end{figure}

The membership matrix $H_g$ for each network represents the
community structure of each network, and the features of $H$ can be
employed to describe the dynamic structure of these networks. For
each run, we can define the following index $S$ for vertex $i$ as
the ratio between the second maximal value and the maximal value of
row $i$ of $H$. The ratio is a positive value less than one. In
reality there is no rigid threshold for significant $S$-score due to
the diversity of networks, but we can select top ones based on the
popular $Z$-score (i.e., $Z=\frac{S-\mu(S)}{\sigma(S)}$, where
$\mu(S)$ is the mean of $S$ and $\sigma(S)$ is the standard
deviation of $S$). By removing the active dynamic vertices according
to this index, we can define the stable co-communities of these
networks.
\begin{figure}[t]
\begin{center}
\includegraphics[width=0.46\textwidth]{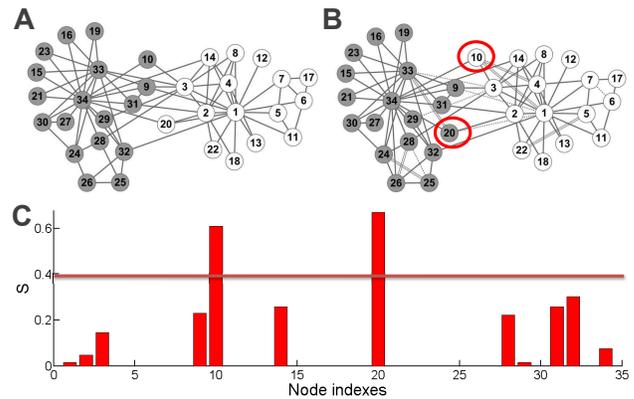}
\caption{(A) The original karate club network. (B) The artificial
evolving network with 12 links' difference compared with the network
in (A). (C) The dynamic index shows the dynamic properties of
vertices.} \label{figurecurve}
\end{center}
\end{figure}

We first test the proposed method using a pair simulated toy
networks representing a time-varying system under two time points
with 16 links' difference (Figure 1A and B). In the system, there
are three clear communities, however, in the two conditions, the
links of some vertices have changed due to some perturbation. We aim
to identify these communities, and at the same time, uncover those
link dynamics that can affect the community structure. We note that
the link dynamics happened within and between communities. The
dynamics happened within a community doesn't affect the community
structure, while that between communities can affect it. For
example, the absence of links (15,11) and (15,20) and the emerging
links (15,28) and (15,26) make the vertex 15 move to another
community. Our method can not only well identify the community
structure, but also can accurately distinguish the link dynamics
that affect the community structure (Figure 2C).

We next apply our method to the karate club network and its variants
with 12 links' difference compared with the original one. The
original karate club network was constructed based on the observed
social interactions between members of a karate club, in which, a
dispute arose and the club split into two clubs. We assumed there
are some changes upon the members' relationship as shown in Figure
3B. Our method can well identify the core communities which
corresponds to the two real sub-clubs (Figure 3A and B). At the same
time, we can uncover the vertices whose link dynamics can affect the
community structure. For example, the links of vertices 10 and
vertices 20 have great difference, and the two vertices are located
at the boundary of two communities. These two nodes have evolved
into opposite communities which can well be reflected by the measure
$S$ (Figure 3C).

\begin{figure}[t]
\begin{center}
\includegraphics[width=0.46\textwidth]{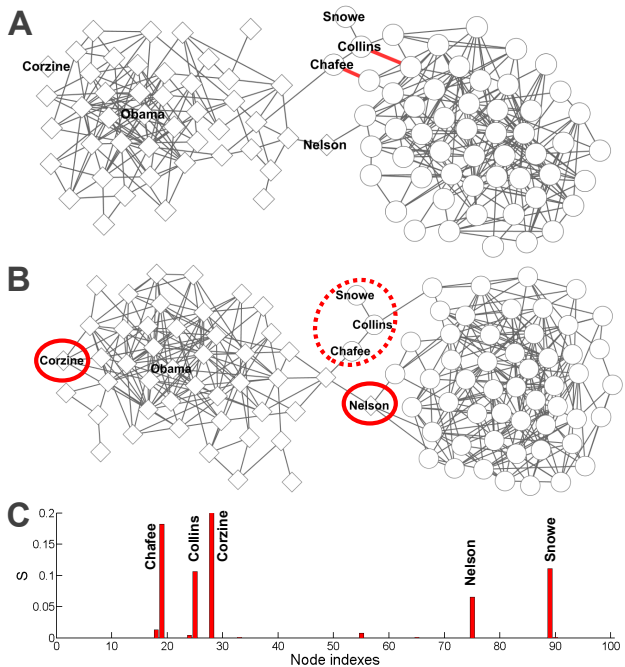}
\caption{(A) The U.S. Senate networks at different time points: (A)
$t=1$, (B) $t=5$, and five vertices show distinct dynamic
characteristics. (C) The dynamic indexes show the dynamic properties
of vertices. Vertex shape show the two political paries: square
means Democrat and circle means Republican. } \label{figurecurve}
\end{center}
\end{figure}
We further apply our method to the set of time-varying networks
consisting of 100 vertices (senators), and 8 time points (i.e., 8
time-varying networks) corresponding to 3-month epochs starting on
Jan 1st 2005 and ending on Dec 31st 2006. The network data were
created using the method developed by Kolar \emph{et al.}
\cite{Kolar2010} based on the United States 109th Congress voting
records and analyzed in Ho \emph{et al.} \cite{Ho2011}. An edge
between two senators in such network indicates that their votes were
mostly similar during that particular epoch. We observed that two
successional networks have relatively small changes. As an example,
we show the networks ($t=1$ and $t=5$) and identify the co-community
among them (Figure 4A and B). Our method can well identify the two
co-communities which perfectly capture party affiliations -
Republican senators are almost always in community 1, while
Democratic senators are almost always in community 2. More
interestingly, we can also identify the dynamic changing of some
vertices which reflect the changes of political opinions of some
senators (Figure 4C). For example, the votes of Democrat Nelson were
unaligned with either Democrats or Republicans at $t=1$, while his
votes were gradually shifting towards Republican which can be found
by the index.

In this report, we investigate the common community structure in
time-varying networks. Rather than treating each slice of a series
of time-varying networks independently, we consider them
simultaneously by defining a common community structure among them.
We have proposed a new framework for recovering the common community
structure and exploring the dynamic changes in these networks by
solving an elaborate mathematical programming problem via existing
decomposition techniques. We have applied the method to both real
and simulated networks, demonstrating that it is able to recover
known co-community structure and reveal dynamic changes among them.
The nondeterministic characteristic of the method allows it for the
selection of number of communities and quantification of the
stability of the community structure. We should note that our
framework can shed lights on the situation that dramatic changes
appear in time-varying networks. Specifically, by applying our
method on each network respectively, we can detect the community
structure of the two networks. And by calculating the consistency of
the two community structure with a measure like normalized mutual
information (NMI) index, we can see how similar the community
structure are in the two networks.

In summary, the main purpose of this report is to propose the new
concept and theoretical framework to analyze the common community
structure of multiple slices of a series of time-varying networks
which shed lights on the network's dynamics and stability. Hope it
can become a promising method to analyze real-world networks. We
need to point out that the adjacency matrix $A$ used in this
framework can be replaced by some \emph{similarity} matrix based on
the connectivity like kernel matrix.


This work was partially supported by the National Natural Science
Foundation of China, No. 11001256, 11131009, 60873205, the `Special
Presidential Prize - Scientific Research Foundation of the CAS, and
the Special Foundation of President of AMSS at CAS for `Chen
Jing-Run' Future Star Program (to S.Z.). The authors thank Professor
Eric P. Xing for providing the network data.

\subsection{Appendix} 
\begin{figure}[h]
\begin{center}
\includegraphics[width=0.44\textwidth]{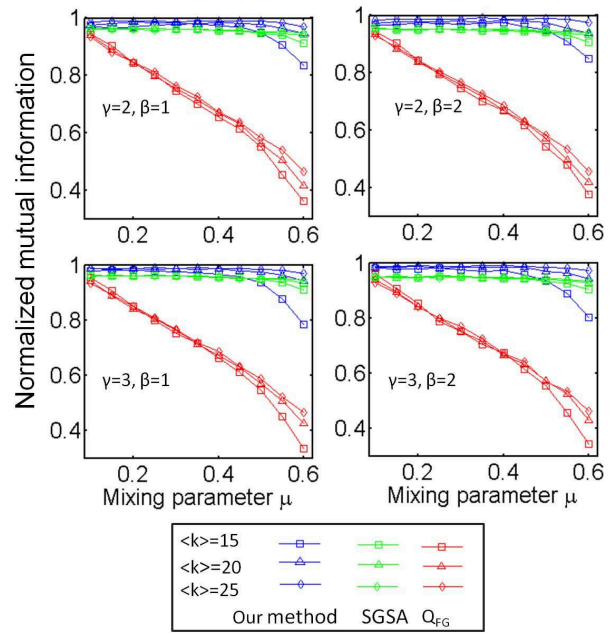}
\caption{Tests of our method on single network using the benchmark
suggested in Lancichinetti \emph{et al} (2008). We also compared it
with two modularity optimization algorithms: the fast greedy
modularity optimization method ($Q_{FG}$) (Phys. Rev. E 2004, 70,
066111) and the spin-glass model and simulated annealing method
(SGSA) (Phys. Rev. E 2006, 74, 016110). Each point corresponds to an
average over 25 network realizations. Detailed parameter settings of
the simulated networks can be seen in Lancichinetti \emph{et al}
(2008). } \label{fig.nmi}
\end{center}
\end{figure}

We have applied the reduced formulation onto
simulated networks with multiple trials. The networks have been
simulated based on the principle suggested in Lancichinetti \emph{et
al} (Phys Rev E 2004, 78, 046110). We found that our method can
obtain reasonable results for many different simulation settings
assessed with normalized mutual information (NMI) index (Figure
\ref{fig.nmi}). We also compared it with other typical community
methods which have shown our method has competitive performance with
them. These analyses partially show that our criterion for multiple
networks is reasonable. \\

\begin{figure}[h]
\begin{center}
\includegraphics[width=0.38\textwidth]{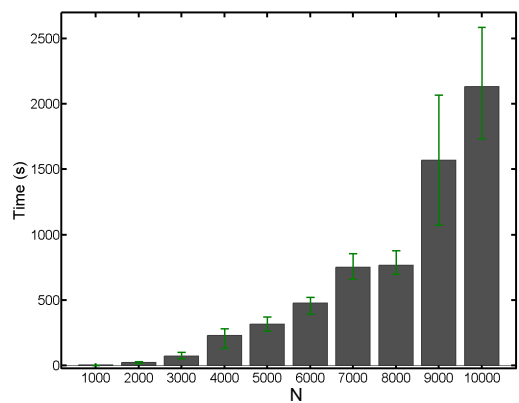}
\caption{The computation time (in seconds) with network size of
n=1000 to 10000. Each bar corresponds to an average over 25 network
realizations. } \label{fig.time}
\end{center}
\end{figure}

The computational efficiency of the proposed method can also be seen
in the simulation study where we have applied the reduced
formulation onto a single network with 10000 nodes. Both of the
theoretic and experimental analyses have shown that our method can
scale well (Figure \ref{fig.time}). \\

[A1] Clauset, A., Newman, M.E.J. and Moore, C. (2004)
Phys. Rev. E 70, 066111.

[A2] Reichardt, J. and Bornholdt, S. (2006) 
Phys. Rev. E, 74, 016110.

\begin{thebibliography}{40}
\bibitem{DT} Note that each element of $H$ in Eq.(4) is an indepndent
variable, so this problem can be decomposed into $N\times K$ one
dimensional subproblem which can be expressed as minimization of
$\lambda_1 \sum_{g=1}^2 |(H_g)_{ij}-(H)_{ij}| + \lambda_2
|(H)_{ij}|$. The optimal solution of this one-dimentional subsection
function subproblem can be easily obtained by considering the value
of each interval.
\end{thebibliography}
\end{document}